\documentclass[11pt,twoside]{article}
\usepackage{asp2004}
\usepackage{psfig}
\usepackage{epsf}
\usepackage{graphics}
\usepackage{lscape}
\def\edcomment#1{\iffalse\marginpar{\raggedright\sl#1\/}\else\relax\fi}
\reversemarginpar

\newcommand{\be}{\begin{equation}}
\newcommand{\ee}{\end{equation}}

 \newcommand{\eqa}{\begin{eqnarray}}
\newcommand{\eeq}{\end{eqnarray}}  
\newcommand{\eqsto}[2]{Eqs. (\ref{#1}) to (\ref{#2})} 

\begin{document}
\title{Non Kolmogorov-like Turbulence in the Local Interstellar Medium}
\author{Dastgeer Shaikh}
\affil{Institute of Geophysics and Planetary Physics,\\
University of California, Riverside, CA 92521. USA.}

\begin{abstract}
We develop a self-consistent model of turbulence in a local
interstellar medium (ISM).  The model describes a partially ionized
magnetofluid ISM in which a neutral hydrogen fluid interacts with a
plasma dominantly through a charge exchange. The ISM turbulent
correlation scales in our model are much bigger than the shock
characteristic length-scales. Unlike small length-scale linear
collisional dissipation in the fluid, the charge exchange processes
can be effective unpredictably on a variety of ISM length-scales
depending upon the neutral and plasma densities, the charge exchange
cross section and the characteristic length scales. We find, from
scaling arguments that the charge exchange interactions modify
spectral transfer associated with large-scale energy containing
eddies. Consequently, the ISM turbulent cascade are steeper than those
predicted by Kolmogorov's phenomenology.
\end{abstract}

\vspace{-0.5cm}

\section{Introduction and Model Equations}
Small scale turbulence in the local interstellar medium (ISM) is a
largely unexplored field given the complexity associated with the ISM
turbulent processes \citep{dastgeer}. The local ISM comprises of
partially ionized, magnetized plasma protons and almost equal number
of neutral particles.  The plasma and the neutral particles in ISM
interact mutually through charge exchange.  The physics of these
small-scale turbulent motions is far more complex than ever
thought. Not only that it holds the key to our crucial understanding
of the global heliospheric interactions such as nature and
characteristic of heliospheric shocks, heating etc
\citep{zank99,pauls}, it is also increasingly believed to be pivotal
to many puzzles of astrophysics including origin and transport of
cosmic rays, Fermi acceleration, gamma-ray bursts, ISM density spectra
etc. Yet, there exists {\it no} self-consistent simulation model that
unravels multi-component and multiple-scale ISM turbulent phenomena.
A prime goal of this paper is therefore to develop a self-consistent
plasma-neutral ISM turbulence model based on analytic methods and
numerical simulations.

The underlying model is based on the following assumptions.
Fluctuations in the plasma and the neutral fluids are sufficiently
isotropic, homogeneous, thermally equilibrated and turbulent. No mean
magnetic field and velocity flows are present at the outset. There
however may generate local mean flows due to self-consistently excited
nonlinear instabilities. The characteristic turbulent correlation
length-scales are typically smaller than charge-exchange mean free
path lengths in the ISM flows. Nevertheless, they are large enough to
treat any localized shocks as discontinuities. In other words, the
characteristic shock length-scales are relatively small compared to
the ISM turbulent fluctuation length-scales, and finally boundary
conditions are periodic. It should be further noted that the neutrals
coming from the solar wind are not considered because they tend to
anisotropize the distribution functions substantially.  Our model thus
simulates a localized ISM. The fluid model describing nonlinear
turbulent processes in the interstellar medium, in the presence of
charge exchanges forces, can then be cast into plasma density
($\rho_p$), velocity (${\bf U}_p$), magnetic field (${\bf B}$),
pressure ($P_p$), as follows.  \be
\label{mhd:cont}
 \frac{\partial \rho_p}{\partial t} + \nabla \cdot (\rho_p{\bf U}_p)=0,
\ee
\be
\label{mhd:mom}
\rho_p \left(  \frac{\partial }{\partial t} + {\bf U}_p \cdot \nabla \right) {\bf U}_p
= -\nabla P_p + \frac{1}{c} {\bf J} \times {\bf B}+{\bf Q}_M({\bf U}_p,{\bf V}_n),
\ee
\be
\label{mhd:mag}
 \frac{\partial {\bf B}}{\partial t} = \nabla \times ({\bf U}_p \times {\bf B}),
\ee
\be
\label{mhd:en}
 \frac{\partial e}{\partial t} 
+ \nabla \cdot \left( \frac{1}{2}\rho_p U_p^2{\bf U}_p 
+ \frac{\gamma}{\gamma-1}\frac{P_p}{\rho_p}\rho_p{\bf U}_p
+ \frac{c}{4\pi}{\bf E} \times {\bf B}  \right)
=Q_E({\bf U}_p,{\bf V}_n).
\ee
where $e=1/2\rho_p U_p^2 + P_p/(\gamma-1)+B^2/8\pi$.
The above set of plasma equations is coupled self-consistently to the
ISM neutral density ($\rho_n$), velocity (${\bf V}_n$) and pressure
($P_n$) through a set of hydrodynamic fluid equations as below.
\be
\label{hd:cont}
 \frac{\partial \rho_n}{\partial t} + \nabla \cdot (\rho_n{\bf V}_n)=0,
\ee
\be
\label{hd:mom}
\rho_n \left(  \frac{\partial }{\partial t} + {\bf V}_n \cdot \nabla \right) {\bf V}_n
= -\nabla P_n + {\bf Q}_M({\bf V}_n,{\bf U}_p),
\ee
\be
\label{hd:en}
 \frac{\partial}{\partial t} \left( \frac{1}{2}\rho_n V_n^2 +
 \frac{P_n}{\gamma-1} \right) + \nabla \cdot \left( \frac{1}{2}\rho_n
 V_n^2{\bf V}_n + \frac{\gamma}{\gamma-1}\frac{P_n}{\rho_n}\rho_n{\bf
   V}_n \right) =Q_E({\bf V}_n,{\bf U}_p).  
\ee 
Equations (\ref{mhd:cont}) to (\ref{hd:en}) form an entirely
self-consistent description of the coupled ISM plasma-neutral
turbulent fluid.  The charge-exchange momentum sources in the plasma
and the neutral fluids are described respectively by terms ${\bf
  Q}_M({\bf U}_p,{\bf V}_n)$ and ${\bf Q}_M({\bf V}_n,{\bf U}_p)$.
Similarly, charge exchange energy sources in the plasma and the
neutral fluids are given by $Q_E({\bf U}_p,{\bf V}_n)$ and $Q_E({\bf
  V}_n,{\bf U}_p)$ respectively.  In the absence of charge exchange
interactions, the plasma and the neutral fluid are de-coupled
trivially and behave as ideal fluids.  While the charge-exchange
interactions modify instantaneous momentum and the energy of plasma
and the neutral fluids, they tend to conserve density in both the
fluids. Nonetheless, the volume integrated energy ($\int Q_E({\bf
  U}_p,{\bf V}_n) d{\bf v}= - \int Q_E({\bf V}_n,{\bf U}_p) d{\bf v} $) and the
density ($\int \rho_p d{\bf v}=const, \int \rho_n d{\bf v}=const$) of
the entire coupled system will remain conserved in the absence of
external sources/sinks.

The ISM turbulence model is normalized using the typical ISM
scale-length ($\ell_0$), density ($\rho_0$) and velocity ($v_0$). The
normalized plasma density, velocity, energy and the magnetic field are
respectively; $\bar{\rho}_p = \rho_p/\rho_0, \bar{\bf U}_p={\bf
U}_p/v_0, \bar{P}_p=P_p/\rho_0v_0^2, \bar{\bf B}={\bf
B}/v_0\sqrt{\rho_0}$. The corresponding neutral fluid quantities are
$\bar{\rho}_n = \rho_n/\rho_0, \bar{\bf U}_n={\bf U}_n/v_0,
\bar{P}_n=P_n/\rho_0v_0^2$. The momentum and the energy
charge-exchange forces, in the normalized form, are respectively
$\bar{\bf Q}_m={\bf Q}_m \ell_0/\rho_0v_0^2, \bar{Q}_e=Q_e
\ell_0/\rho_0v_0^3$. The non-dimensional temporal and spatial
length-scales are $\bar{t}=tv_0/\ell_0, \bar{\bf x}={\bf
x}/\ell_0$.   The charge-exchange cross-section parameter
($\sigma$), not appeared directly in the above set of equations, is
normalized as $\bar{\sigma}=n_0 \ell_0 \sigma$.  We define an
intrinsic mode in ISM turbulence, the charge exchange mode $k_{ce}\sim
(n_0\ell_0)^{1/2}$. The latter is different from the characteristic
turbulent mode $k$, and is excited naturally when plasma and neutral
fluids are coupled in the ISM by charge exchange forces [see
\citet{dastgeer} for more detail].

\section{ Nonlinear Simulation}

Two-dimensional (2D) nonlinear spectral fluid code is developed to
numerically integrate \eqsto{mhd:cont}{hd:en}. The 2D simulations are
not only computationally simpler and less expensive (compared with the
full 3D), they offer significantly higher resolutions even on moderate
size small cluster machines like Beowulf.  The spatial discretization
in our code uses discrete Fourier representation of turbulent
fluctuations based on a spectral method, $\tilde{f}({\bf
  x,t})=\sum_{\bf k} f({\bf k},t) e^{-i{\bf k} \cdot {\bf x}}$, where
${\bf k}=k_x \hat{e}_x + k_y \hat{e}_y$ is a two-dimensional wave
vector. The nonlinear deconvolution of Fourier modes is performed by
computing nonlinear triad interaction $\tilde{f}({\bf x,t})
\tilde{g}({\bf x,t}) = \sum_{{\bf k}={\bf k}'-{\bf k}''} f({\bf
  k}',t)g({\bf k}'',t) \delta({\bf k}'-{\bf k}'')$ which survive for
only those mode coupling interactions which satisfy Fourier diad
constraint ${\bf k}={\bf k}'-{\bf k}''$. These nonlinear interactions
in Fourier space conserve rugged invariants of system of
\eqsto{mhd:cont}{hd:en}.  The temporal integration is performed by
Runge Kutta 4 method. The fluctuations in plasma and neutral are
initialized isotropically (no mean fields are assumed) with random
phases and with identical amplitudes in the Fourier space.  This
algorithm ensures conservation of total energy and mean fluid density
per unit time in the absence of charge exchange and external random
forcing.  Additionally, it satifies the condition of incompressibility
associated with the magnetic field, i.e. $\int |\nabla \cdot {\bf
  B}|^2 d{\bf v} \approx 10^{-15}$ at each time step. We make use of
an artificial scalar potential to achieve $\nabla \cdot {\bf B}=0$ as
described in \citet{nick}.  Numerical resolution is $512^2$ for the
mode structures and $1024^2$ for the spectrum calculations.  Our code
is massively parallelized using Message Passing Interface (MPI)
libraries to facilitate higher resolution.  While the ISM turbulence
code is evolved with time steps resolved self-consistently by the
coupled fluid motions, the nonlinear interaction time scales
associated with the plasma $1/{\bf k} \cdot {\bf U}_p({\bf k})$ and
the neutral $1/{\bf k} \cdot {\bf V}_n({\bf k})$ fluids can obviously
be disparate. Accordingly, turbulent transport of energy in the plasma
and the neutral ISM fluids takes place distinctively on separate time
scales.
\begin{figure}[!t]
\begin{center}
\psfig{file=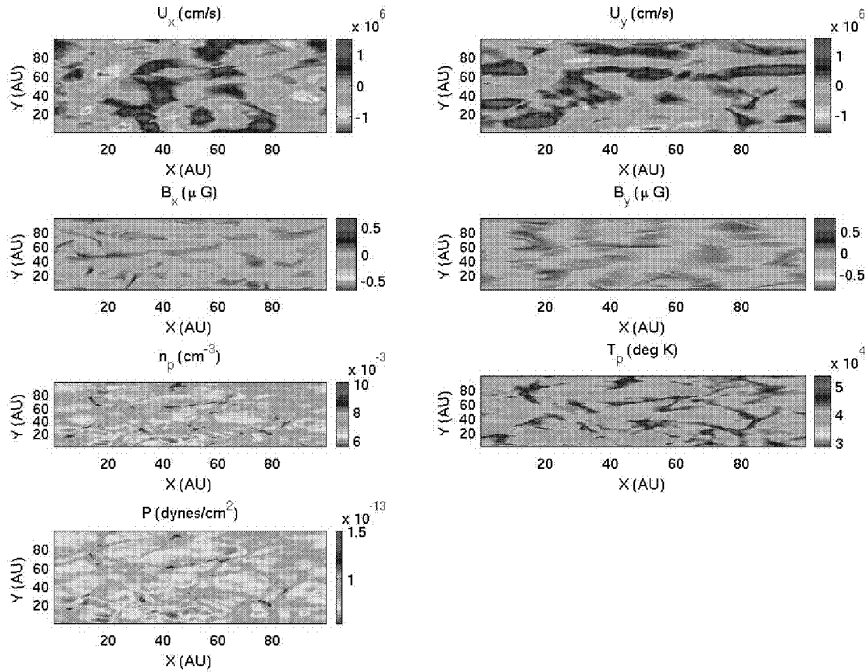,width=4.5in}
\end{center}
  \caption{Self-consistent simulations of turbulence in the partially
    ionized ISM. Figure shows evolution of the plasma density,
    velocity, magnetic field, temperature and pressure. Numerical resolution is
$512^2$.}
\end{figure}
Small scale turbulent fluctuations in ISM evolve under the action of
nonlinear interactions as well as charge exchange sources. Energy
cascades amongst turbulent eddies of various scale sizes and between
the plasma and the neutral fluids. Because of the discrepant nonlinear
time scales associated with the plasma and neutral fluids, mode
structures can be different in the two fluids when they are evolved
together and in isolation (i.e. decoupled). When the plasma and
neutral components are decoupled, the plasma fluid evolves in
accordance with the ideal MHD where current sheets are typically
formed in the magnetic field structures.  In the presence of charge
exchange, the two fluids evolve in a self-consistent manner and modify
the dynamical properties of both the fluids. One of the most
noticeable features to emerge from the coupled evolution is that the
spectral cascade rates are enhanced substantially in contrast to that
of the decoupled case. The rapid spectral transfer of energy amongst
various Fourier modes in the coupled plasma neutral gas leads to
smearing off the current sheets in the plasma on a shorter time
scales. The latter evolves on a relatively long time scales and the
sheets are formed eventually in the magnetic field. The turbulent
equipartition is set up in the plasma and the neutral fluid
modes. This is shown in Fig 1 where all the quantities associated with
the plasma evolution in our simulations are depicted.  Seemingly, the
neutral fluid, under the action of charge exchange sources, tends to
enhance the cascades rates by isotropizing the ISM turbulence on rapid
time scales.  One of the plausible reasons of the fast-cascade is that
the energy transfer rates are enhanced effectively in the coupled ISM
turbulence system, and that they now require a smaller time to mix the
plasma and the neutral fluids.  In any event, the small-scale
sheet-like structures in magnetic field compresses (or pinches) the
plasma density. Hence the density fluctuations develop identically the
(thinner than) sheet-like structures that co-exist with small-scale
turbulent fluctuations in its spectrum as shown in Fig 1. The neutral
fluid, on the other hand, evolves isotropically as stated above by
forming relatively large-scale structures as shown in Fig 2.

\begin{figure}[!t]
\begin{center}
\psfig{file=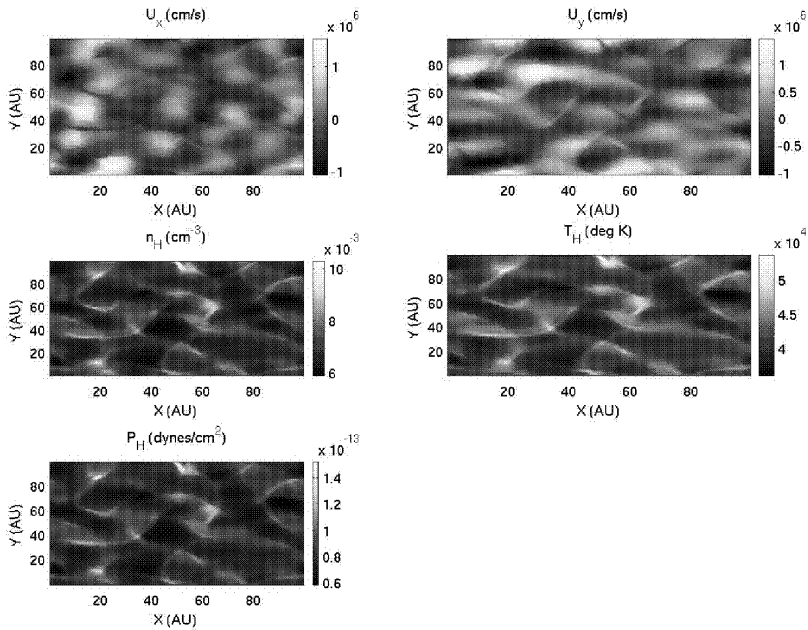,width=4.2in}
\end{center}
  \caption{Evolution of neutral fluid in the coupled plasma neutral
    ISM turbulence. Shown are the neutral density, velocity,
    temperature and pressure fluctuations.}
\end{figure}

\section{Energy Cascade Rates in the Coupled ISM Plasma Neutral Gas}
As mentioned above, the plasma neutral fluid coupling in the ISM
fluctuations introduces charge exchange modes, $k_{ce}$, that are
distinctively different from the characteristic turbulent mode
$k$. Typically, $k_{ce}/k<1$ in the local ISM. Accordingly, the energy
cascade time scales are modified as discussed below.  The nonlinear
interaction time-scale in ordinary turbulence is given by
\[\tau_{nl} \sim \frac{\ell_0}{v_\ell} \sim (kv_k)^{-1},\] 
where $v_k$ or $v_\ell$ is the velocity of turbulent eddies. In the
presence of charge exchange interaction, the ordinary nonlinear
interaction time-scales of fluid turbulence is modified by a factor
$k_{ce}/k$ such that the new nonlinear interaction time-scale of ISM
turbulence is now 
\[\tau_{NL} \sim \frac{k_{ce}}{k} \frac{1}{kv_k}.\] 
On using the fact that $k_{ce}$ is typically smaller than $k$,
i.e. $k_{ce}/k<1$ in the ISM, the new nonlinear time is $k_{ce}/k$
times smaller than the old nonlinear time i.e.  $\tau_{NL} \sim
(k_{ce}/k) \tau_{nl}$. This reduced nonlinear interaction time in ISM
turbulence is likely to enhance turbulent energy cascade rates that
are determined typically by $E_k/\tau_{NL}$, where $E_k$ is energy per
unit mode.  It is beacuse of this reduced interaction time that a
rapid spectral transfer of turbulent modes tends to smear off the
current sheets in the magnetic field fluctuations.

It is interesting to note that the enhanced cascade rates of ISM
turbulent modes tend to steepen the inertial range turbulent spectra
in both plasma and neutral fluids. By extending above phenomenological
analysis, one can deduce exact (analytic) spectral indices of the
inertial range decaying turbulent spectra, as follows. The new
nonlinear interaction time-scale of ISM turbulence can be rearranged
as $ \tau_{NL} \sim
k_{ce}v_k /(kv_k)(1/kv_k)\sim ({\tau_{nl}^2}/{\tau_{ce}})$,
where $\tau_{ce}\sim (k_{ce} v_k)^{-1}$ represents charge exchange
time scale. The energy dissipation rate associated with the coupled
ISM plasma-neutral system can be determined from $\varepsilon \sim
E_k/\tau_{NL}$ relation, which leads to 
$\varepsilon \sim {v_k^2}/({k_{ce}/k^2 v_k})\sim {k^2 v_k^3}/{k_{ce}}$.
According to the Kolmogorov theory, the spectral cascades are local in
$k$-space and the inertial range energy spectrum depends upon the
energy dissipation rates and the characteristic turbulent modes, such
that $E_k \sim \varepsilon^{\gamma} k^{\beta}$.  Upon substitution of
above quantities and equating the power of identical bases, one
obtains
\[E_k \sim \varepsilon^{2/3} k^{-7/3}\] 
plasma spectrum.  Similar arguments in the context of neutral fluids,
when coupled with the plasma fluid in ISM, lead to the energy
dissipation rates $\varepsilon \sim k^2 v_k^2/({k_{ce}/k^2
  v_k})$. This further yields the forward cascade (neutral) energy
spectrum 
\[E_k \sim \varepsilon^{2/3} k^{-11/3}.\]

\section{Conclusion}
In conclusion, one of the most important points to emerge from our
studies is that the charge exchange modes modify the ISM turbulence
cascades dramatically by expediting nonlinear interaction time-scales.
consequently, the energy cascade rates are enhanced by isotropizing
the ISM turbulence on rapid time scales.  This tends to modify the
characteristics of ISM turbulence which can be significantly different
from the Kolmogorov phenomenology of fully developed turbulence. As
shown above, the charge exchange interactions lead to a steeper power
spectrum.  Although turbulence in 2D and 3D possesses distinct
spectral features characterized essentially by the number of the
inviscid quadratic invariants, our 2D simulations provide a hint that
the ISM turbulent spectra in 3D case will exhibit a non
Kolmogorov-like characteristic.  It is to be noted that the present
model does not consider driving mechanism, hence turbulence is freely
decaying.  Driven turbulence, such as due to large scale forcing,
supernova explotion and CME ejection may force turbulence at larger
scales. This can modify the cascade dynamics in a manner usually
described by dual cascade process.

\end{document}